\documentstyle[prb,preprint,aps]{revtex}

\tighten
\begin{document}
\author{Alejandro Cabo*, Francisco Claro and Alejandro \ P\'{e}rez}
\address{Facultad de F\'{i}sica, Pontificia Universidad\\
Cat\'{o}lica de Chile, Vicu\~{n}a Mackenna 4860, 6904411, Macul,\\
Santiago de Chile}
\title{The Hartree-Fock state for the 2DEG at $\nu =\frac 12$ revisited: analytic
solution, dynamics and correlation energy.}
\date{May 5, 2001}
\maketitle

\begin{abstract}
The CDW Hartree-Fock state at half filling and half electron per unit cell
is examined. Firstly, an exact solution in terms of Bloch-like states is
presented. Using this solution we discuss the dynamics near half filling and
show the mass to diverge logarithmically as this filling is approached. We
also show how a uniform density state may be constructed from a linear
combination of two degenerate solutions. Finally we show the second order
correction to the energy to be an order of magnitude larger than that for
competing CDW solutions with one electron per unit cell.

\medskip


\ \bigskip

* On leave of absence from {\it Theoretical Physics Group, Instituto de
Cibern\'etica, Matematica y F\'{i}sica, Calle E, No. 309, Vedado, La Habana,
Cuba. }
\end{abstract}

\section{Introduction}

For almost two decades the fractional quantum Hall effect has been one of
the main focus of interest in condensed matter physics\cite{stormer,laughlin}%
. Much progress has been made towards its understanding, and even general
theories exist today aspiring to be fully coherent descriptions of the
underlying physics\cite{book}. These theories do not rest on an actual
solution of the basic quantum mechanical equations of motion, however, but
are rather cast from ansatz wave functions exhibiting a large overlap with
accurate numerical solutions of such equations for a few particles. A more
basic theory, although highly desirable, is very difficult to attain because
electron-electron interactions and correlations are at the core of the
effect.

A first step in a perturbative approach was developed in the early years by
Yoshioka and Lee, who constructed a mean field Hartree-Fock theory for the
spin polarized case (HFT1)\cite{yoshioka lee}. It received little attention,
however, because it failed to provide the empirical selection rule that
distinguishes even from odd denominator filling fractions, which
characterizes the effect in an essential way. It was later shown by one of
us, that if the mean field solution is constructed in a slightly different
way, such distinction arises naturally, giving the same gap structure at the
various filling fractions as experiment, and producing the proper step like
Hall conductivity dependence with magnetic field (HFT2)\cite{claro1,claro2}.
It predicts a gap at every odd denominator fraction, and a metallic state at
all even denominators, within a band spectrum whose fine structure at or
near the Fermi energy scales with the denominator in the fraction. Energies
were too high, however, even compared to the state proposed by Yoshioka and
Lee. One potential problem of Hartree-Fock spin polarized states in the
lowest Landau level is that they necessarily have non uniform electron
density at fractional filling \cite{claro2}, and, if space fluctuations are
severe, possibly pin the many-body electronic state to the underlying
impurities, an effect for which there appears to be no experimental evidence
except perhaps at fillings below 1/7. However, it should be also stressed
that a crystalline ordered state (''Hall Crystal'') can be fully compatible
with a quantized Hall conductivity thanks to its magnetic field dependent
crystal parameters\cite{halperin}.

Simple charge density wave (CDW) mean field states are found assuming they
form a periodic lattice of rectangular or triangular geometry. The size of
the unit cell is an additional degree of freedom characterized by the
quantity $\gamma $, the number of electrons per unit cell. The main
difference between the mean field theories described above is that HFT1
assumes one electron per unit cell, that is, $\gamma =1$, while HFT2 assumes 
$\gamma =1/2$ or some other fraction of denominator $2^k$, with $k$ an
integer. Keeping the geometry fixed and changing $\gamma $ yields local
minima at $\gamma =1/2,1$. Although neither show energies near the one
corresponding to the true ground state, one may ask which one is a better
perturbative precursor of the ground state. HFT1 gives a lower energy, but
HFT2 provides the basic selection rule distinguishing even and odd
denominators as required by experiment, and has less pronounced charge
density fluctuations.

In this work we present an analytic solution for the $\nu =1/2,$ $\gamma
=1/2 $ state. In particular, we address the question of sensitivity to
perturbation theory in both theories. Within HFT1, Yoshioka and Lee obtained
a correction to second order of about 0.002 in units of $\frac{e^{2}}{r_{o}}$%
, where $r_{o}$ is the magnetic length \cite{yoshioka lee}. Here we show the
same correction to be an order of magnitude larger for HFT2. This result
confirms the suggestion made in a former work by one of us about that this
latter state is more sensitive to correlations than HFT1\cite{cabo1}. Our
solution is constructed in terms of Bloch-like running waves which solve the
Hartree Fock problem exactly \cite{cabo0} and form a complete orthonormal
set, save for a single point in the Brillouin zone \cite{wannier}. This
remarkable property may be unique to filling $1/2$ and $\gamma =1/2$, since
then exactly one flux quanta traverses the unit cell of the underlying CDW .
Having analytic solutions allows for a study of the dynamics at the Fermi
surface. We find that the cyclotron effective mass diverges logarithmically
as half filling is approached, in agreement with RPA estimates \cite{halp}
and with some experimental data \cite{du}.

In Section 2 the single particle solutions and their self-energies are
given. Close expressions in terms of products of elliptic theta functions
are presented, and the order parameter is explicitly given. We also show
that degenerate Hartree-Fock solutions may be superposed to form a state of
uniform charge density. In Sec. 3 a semiclassical analysis of the response
to a small additional field is given, and the resulting cyclotron mass
discussed. Section 4 is devoted to the evaluation of the second order
correction to the energy. A short review and discussion of our results is
given in the Summary. In the appendices we derive some symmetry properties
of our solutions and show that filling the rotated central square of the
Brillouin zone indeed yields the lowest self-consistent energy solution.

\section{ Analytic single particle solutions and order parameters}

We consider $N$ electrons constrained to move in a plane of area $S$ under a
perpendicular magnetic field ${\bf B}$. As shown by Wannier\cite{wannier},
to study this system one can construct an orthogonal set of single particle
Bloch-like states from the zero angular momentum eigenfunction in the lowest
Landau level

\begin{equation}
\phi ({\bf x})=\frac 1{\sqrt{2\pi }r_o}\exp (-\frac{x^2}{4r_o^2}),
\end{equation}
in the form\cite{ferrari,cabo0,cabo1},

\begin{equation}
\varphi _k({\bf x})=\frac 1{N_{{\bf k}}}\sum_{{\bf \ell }}\left( -1\right)
^{\ell _1\ell _2}\exp (i\ {\bf k}.{\bf \ell )\ T_{{\ell }}\ \phi (x).}
\end{equation}
Eigenstates are labeled by a wave vector ${\bf {k}={p}}/\hbar $ where ${\bf {%
p}}$ is the quasi-momentum, and the sum runs over all integers $\ell _1,\ell
_2$ defining a planar square lattice $L$, with ${\bf {\ell }}=a(\ell _1,\ell
_2,0),\ a=2\pi r_o^2$ and $r_o=\sqrt{\frac{\hbar c}{|e|B}}.$The magnetic
translation operator ${\bf {T_\ell }}$ acting on any function $f$ introduces
a phase: 
\begin{equation}
{\bf {T_\ell }\ }f({\bf x}){\bf =\exp (\frac{ie}{\hbar c}A({\ell }).x)\ }f(%
{\bf x-{\ell }}){\bf ,}
\end{equation}
where the vector potential is assumed in the axial gauge ${\bf A}({\bf x})=%
\frac B2(-x_2,x_1,0)$ and the electron charge $e$ is taken with its negative
sign.

We want to study the case in which one flux quanta traverses each lattice
cell. Then, there is one state per plaquette in the lowest Landau level and
at filling one half the charge in each cell is just half the electron
charge, and thus, $\gamma =1/2$. The wave-function (2) and the normalizing
factor $N_{{\bf {k}}}$ can then be expressed in terms of elliptic theta
functions thanks to their simple properties under special shifts of the
complex arguments in their quasi-periods. They are given, respectively, by

\begin{eqnarray}
\varphi _{{\bf k}}({\bf x}) &=&\frac{1}{\sqrt{2\pi }r_{o}N_{k}}\exp (-\frac{%
x^{2}}{4r_{o}^{2}})\ \left( \Theta _{3}(\frac{k_{1}^{*}a}{\pi }\mid -\frac{1%
}{\tau })\Theta _{3}(\frac{k_{2}^{*}a}{2\pi }\mid \tau )+\Theta _{2}(\frac{%
k_{1}^{*}a}{\pi }\mid -\frac{1}{\tau })\Theta _{4}(\frac{k_{2}^{*}a}{2\pi }%
\mid \tau )\right) ,  \nonumber \\
N_{k}^{2} &=&N_{\phi _{0}}(\Theta _{3}(\frac{k_{1}a}{\pi }\mid -\frac{1}{%
\tau })\Theta _{3}(\frac{k_{2}a}{2\pi }\mid \tau )+\Theta _{2}(\frac{k_{1}a}{%
\pi }\mid -\frac{1}{\tau })\Theta _{4}(\frac{k_{2}a}{2\pi }\mid \tau )), 
\nonumber
\end{eqnarray}
where $\tau =\frac{i}{2}$ . The components of the wave vectors appearing in
these expressions are related through ${\bf k}{^{*}}={\bf k}+({\bf n}\times 
{\bf x}-i\ {\bf x})/2r_{0}^{2}$, where ${\bf n}=(0,0,1)$ is the unit vector
normal to the plane. The orthonormal set is well defined except at the
single point ${\bf k}=(\pi /a,\pi /a,0)$, where the norm vanishes\cite
{wannier}. In what follows we shall ignore this singular point.

As may be easily verified, the functions $\varphi _{{\bf k}}\ $ satisfy the
eigenvalue equation

\begin{eqnarray}
T_{{\bf \ell }}\ \varphi _{{\bf k}}({\bf x}) &=&\lambda _{{\bf k}}({\bf \ell 
})\ \varphi _{{\bf k}}({\bf x}), \\
\lambda _{{\bf k}}({\bf \ell }) &=&\left( -1\right) ^{\ell _1\ell _2}\exp
(-i\ {\bf k}.{\bf \ell }).\   \nonumber
\end{eqnarray}
As was shown earlier they are exact solutions of the Hartree-Fock (HF)
single particle Schrodinger equation associated to an arbitrary Slater
determinant formed by selecting an also arbitrary group as filled states 
\cite{cabo0,cabo2}. This occurs because the HF single particle Hamiltonian
commutes with all translations leaving $L$ invariant \cite{cabo0}. Since the
functions (2) are common eigenfunctions of the commuting magnetic
translations leaving invariant the lattice $L$ and the set of eigenvalues
(4) uniquely determines them, the HF potential associated with the Slater
determinant can not change those eigenvalues. Therefore $\varphi _{{\bf k}}$
should be an eigenfunction.

The explicit expression for the self-energy of the state $\varphi _{{\bf k}} 
$ has the form, \cite{cabo1}

\begin{equation}
{\Large \epsilon }({\bf k})\ =\frac{S}{2\pi r_{o}^{2}}\sum_{{\bf Q}}\Delta _{%
{\bf k}}({\bf Q})\ \Delta ^{*}({\bf Q})\exp (-\frac{r_{o}^{2}{Q}^{2}}{4}%
)\left( \frac{1-\delta _{{Q},0}}{r_{o}Q}\exp (-\frac{r_{o}^{2}Q^{2}}{4})-%
\sqrt{\frac{\pi }{2}}I_{o}(\frac{r_{o}^{2}Q^{2}}{4})\right) \frac{e^{2}}{%
\varepsilon r_{o}},
\end{equation}
where {${\bf Q}=2\pi (n_{1},n_{2},0)/a$} is the set of reciprocal lattice
vectors and $I_{0}$ is the modified Bessel function. The momentum dependence
of the self-energy is fully contained in the factor $\Delta _{{\bf k}}({\bf Q%
})=2\pi r_{o}^{2}\lambda _{{\bf k}}({\bf x}^{*})/S$, where ${\bf x}{^{*}}%
=r_{o}^{2}\ {\bf n}\times {\bf Q}$. The order parameter is in turn given by%
\cite{cabo1}, 
\begin{eqnarray}
\Delta ({\bf Q}) &=&\frac{2\pi r_{o}^{2}}{S}\sum_{{\bf k}\in F}\lambda _{%
{\bf k}}({\bf \ell }) \\
&=&\frac{2\pi r_{o}^{2}}{S}\left( -1\right) ^{n_{1}n_{2}}\sum_{{\bf k}\in
F}\exp (-i\ {\bf k}.{\bf x}^{*}),  \nonumber
\end{eqnarray}
where the sum is over all filled states in the Brillouin zone $B$, the set
of which we call $F$ $\subset B\;$. From general symmetry properties it is
shown in Appendix B that an energy minimum is obtained among all possible
Slater determinants by filling states inside the square bounded by the
constant energy lines $\pm k_{x}\pm k_{y}=\pi /a$, which we take as the
Fermi surface \cite{cabo2}. The number of states in this square is just half
the total, as required, since we are studying the half filling case. Note
that the energy function is continuous across these lines so there is no gap
in the single particle spectrum at the Fermi energy. Turning the sum into an
integral one then gets,

\begin{eqnarray}
\Delta ({\bf Q}) &=&\left( -1\right) ^{n_1n_2}a^2\int_{{p}\in F}\frac{d{\bf k%
}}{(2\pi )^2}\exp (-i\ {\bf k}.{\bf x}{^{*}})\  \\
&=&\left( \frac{\delta _{n_1,\ n_2}}2-\frac 2{\pi ^2}\lim_{\epsilon
\rightarrow 0}\frac{\sin [(n_1-n_2)\frac \pi 2]\sin [(n_1+n_2)\frac \pi 2]}{%
n_1^2-n_2^2+\epsilon }\right) .  \nonumber
\end{eqnarray}
Notice that these quantities vanish if $n_1,n_2$ have the same parity, save
at the origin. Using the above expressions the total energy per electron may
be easily computed, to obtain $\epsilon _{\frac 12}=-0.394\frac{e^2}{%
\varepsilon r_o}$, a value somewhat above the result for $\gamma =1$, $%
\epsilon _1=-0.443\frac{e^2}{\varepsilon r_o}$ \cite{kura }.

The charge density may also be obtained from the above expressions since
their Fourier coefficients and (7) are related through $\Delta ({\bf Q}%
)=2\pi r_{o}^{2}e^{\frac{r_{o}^{2}Q^{2}}{4}}\rho ({\bf Q})$, yielding a
space fluctuation of a mere $17\%$ about the average $n_{0}=N/S$, while the $%
\gamma =1$ state electron density varies between $0.1n_{0}$ and $1.85n_{0}$,
representing a fluctuation of close to 90\% about the average.\cite{kura}

Finally in this section we would like to underline the curious point that
using our Hartree-Fock solutions it is possible to construct a state of
uniform charge density. To see this we consider the function $\Psi =(\Phi
^{HF}+\widetilde{\Phi }^{HF})/\sqrt{2},$ where as before $\Phi ^{HF}$ is the
Slater determinant formed with all states in $F$, while $\widetilde{\Phi }%
^{HF}$ is the Slater determinant of all states in $B-F$, the complement of $%
F $ in the Brillouin zone $B$. Both regions are separated by the Fermi
surface, the square $\pm k_{x}a\pm k_{y}a=\pi $. Because all single-particle
states are orthogonal to each other the crossed term in the charge density
vanishes and one has{\LARGE \ }$\rho (x)=(\rho ^{HF}(x)+\widetilde{\rho }%
^{HF}(x))/2=\sum_{{\bf k}\in B}|\varphi _{{\bf k}}({\bf x})|^{2}/2$, half
the density of a completely filled Landau level, which equals $(4\pi
r_{o}^{2})^{-1}$and is uniform throughout the sample.

We now show that the above linear combination has the minimal energy. To see
this we first note that $\widetilde{\Phi }^{HF}$ is obtained simply by
shifting all momenta in $\Phi ^{HF}$ by the vector ${\bf \delta }=(1,-1)\pi
/a,$ since in the extended zone scheme this vector maps all states in $F$
onto $B-F$ . However, according to Eq. A.2 such shift may be done with the
aid of a single magnetic translation $T_{{\bf R}}$, with ${\bf R}=r_{o}^{2}\ 
{\bf n}\times {\bf \delta =}(1,1)a/2.$ Thus, both Slater determinants are
related by a single translation in space. But, all the N states forming each
of these Slater determinants are orthogonal with all the ones defining the
other, and also the energy operator is a linear combination of products of
merely four creation or annihilation operators. It therefore follows that
the mean value of the energy operator in the state $\Psi $ is just half what
one obtains adding the mean values of $\Phi ^{HF}$ and $\widetilde{\Phi }%
^{HF}$, and thus the mean energy of the superposition state coincides with
the HF energy.

\section{ Dynamics and mass}

It has been suggested that the effective mass of the charged carriers may
diverge at the Fermi surface, a property that remains controversial.\cite
{halp,will} Taking advantage of our analytic results we examine these
questions. The semiclassical equations of motion may be solved for particles
at the Fermi surface moving in the external field corresponding to half
filling, ${B_{\frac{1}{2}}}$, with the result that if motion starts at ${\bf 
{k}=\frac{\pi }{2a}(}1{\bf ,}1{\bf )}$ then, after a time $t$, the particle
quasi- momentum has reached the point $k_{x}=\frac{2}{a}\arctan [\exp
(-t/T)],$ $ak_{y}=\pi -ak_{x}$, with a time constant of order $\tau =2\hbar
/\epsilon ^{*}$, where $\epsilon ^{*}=\frac{e^{2}}{\varepsilon r_{o}}$. It
thus takes an infinite time to reach a corner of the square Fermi surface,
and the trajectory in real space is a straight line covering just a fraction
of the unit cell. One can thus claim that the Fermi particles behave as if
there was no external field at all, in agreement with previous work.\cite
{halp}

In order to examine the effective mass that this solution provides we
consider the period of a cyclotron orbit when the filling is slightly above
or below $1/2$. For definiteness we assume the later. For simplicity me keep
the lowest order Fourier components in the dispersion relation only, and
assume that the slight change in filling fraction does not perturb
significantly the particle density. The self energy of state ${\bf {k}}$ is
then of the form $\epsilon =\epsilon _{o}-\epsilon _{1}($cos($ak_{x}$)$+$cos(%
$ak_{y}$)$)$, with $\epsilon _{o}=-\frac{1}{2}\sqrt{\frac{\pi }{2}}\epsilon $
and $\epsilon _{1}=0.087\ \epsilon ^{*}$. We study the dynamics governed by
the equations

\[
\hbar \frac{d{\bf {k}}}{dt}=-\frac e{\hbar c}\frac{\partial {\epsilon }}{%
\partial {\bf {k}}}\times {\bf {B.}} 
\]

\noindent For an orbit over the Fermi surface, no longer square in shape,
this equation becomes, for the component $k_y$,

\[
\hbar \frac{dk_y}{dt}=-\frac{a\epsilon _1}{\delta ^2}\sqrt{1-(\eta +\cos
(ak_y))^2}, 
\]

\noindent from whose solution and the constant energy condition the
itinerary of $k_{x}$ may be extracted. Here $\delta ^{2}=\hbar c/eB=2\pi
r_{o}^{2}$, and $\eta =(\epsilon -\epsilon _{o})/\epsilon _{1}$ measures the
departure from the half filling square Fermi surface. Integrating this
equation one obtains for the time it takes for a particle to go around the
energy contour once,

\[
T=\frac{4\hbar }{\pi \epsilon _1}\frac{K(\frac{\eta ^2-4}{\eta ^2})}\eta , 
\]
where $K(u)$ denotes the complete elliptic function of the first kind. A
cyclotron effective mass may be obtained from this expression through the
usual definition $m^{*}=eTB/(2\pi c)$. One obtains the result $m^{*}=8\hbar
^2n_0K(\frac{{\eta ^2-4}}{\eta ^2})/(\pi \epsilon _1\eta )$. Noting that $K(%
\frac{{\eta ^2-4}}{\eta ^2})\sim -0.57\eta $ log $\eta $ as $\eta \to 0$ we
get the final result,

\[
m^{*}=-\frac{16.7\text{ }\hbar ^{2}n_{0}}{\epsilon ^{*}}\text{log (}\eta 
\text{)}, 
\]

\noindent which diverges logarithmically as half filling is approached.

\section{Energy per particle in second order}

We now turn our attention to the second order correction to the energy per
particle for the $\nu =\frac{1}{2}$, $\gamma =\frac{1}{2},$ state. A similar
evaluation for $\gamma =1$ was done by Yoshioka and Lee, obtaining a
correction of the order of 0.5\% of the total result.\cite{yoshioka lee} .
However, as it was underlined in \cite{cabo1}, the increased degree of
overlap of the single electron states associated to the $\gamma =\frac{1}{2}$
wave-functions could change the situation drastically. The study of this
question is the main objective of the present paper.

For evaluating the second order correction the following formula will be
employed \cite{yoshioka lee}

\begin{eqnarray}
E^{(2)} &=&\sum_{i}\langle \Phi ^{HF}\mid (H-H_{F}^{(HF)})\mid \Phi
_{i}\rangle \frac{1}{E^{(HF)}-E_{i}}\langle \Phi _{i}\mid
(H-H_{F}^{(HF)})\mid \Phi ^{HF}\rangle \\
&=&\sum_{i}\frac{\mid \langle \Phi ^{HF}\mid H\mid \Phi _{i}\rangle \mid ^{2}%
}{E^{(HF)}-E_{i}}  \nonumber
\end{eqnarray}
where $\Phi ^{HF}$, $E^{(HF)}$ are the Slater determinant and total
Hartree-Fock energy, respectively, and $H$ is the projection of the exact
Hamiltonian onto the first Landau level. The many particle excited states $%
\Phi _{i}$ are Slater determinants constructed with the Hartree-Fock basis
states \{$\varphi _{{\bf k}}$\} , ${\bf k}\in B.$ It follows that $\langle
\Phi ^{HF}\mid \Phi _{i}\rangle =0$, a property that allowed to write the
last equality in (8). In the second quantized representation the Hamiltonian 
$H$ will have non-vanishing matrix elements linking the HF state and excited
states of the form $\mid \Phi _{i}\rangle =a_{{\bf \eta }}\,a_{{\bf \eta }%
^{\prime }}\,a_{{\bf \xi }}^{+}\,a_{{\bf \xi }^{\prime }}^{+}\mid \Phi
^{HF}\rangle ,$ where $a_{{\bf \xi }}^{+}$ creates an electron of wavevector 
${\bf \xi }$, etc. The index $i$ is a shorthand notation for pairs of filled 
$({\bf \eta },{\bf \eta }^{\prime }\in F)$ and empty (${\bf \xi },{\bf \xi }%
^{\prime }\in B-F$) electron states. The total energies of the excited
states are given by $E_{i}=E^{(HF)}+\epsilon ({\bf \xi })+\epsilon ({\bf \xi 
}^{\prime })-\epsilon ({\bf \eta })-\epsilon ({\bf \eta }^{\prime })$. Then,
the second order correction can be rewritten in the form

\begin{equation}
E^{(2)}=\sum_{({\bf \eta },{\bf \eta }^{\prime })}\sum_{({\bf \xi },{\bf \xi 
}^{\prime })}\frac{\mid \langle \Phi ^{HF}\mid H\,a_{{\bf \eta }}\,a_{{\bf %
\eta }^{\prime }}\,a_{{\bf \xi }}^{+}\,a_{{\bf \xi }^{\prime }}^{+}\mid \Phi
^{HF}\rangle \mid ^{2}}{{\Large \epsilon }({\bf \eta })+{\Large \epsilon }(%
{\bf \eta }^{\prime })-{\Large \epsilon }({\bf \xi })-{\Large \epsilon }(%
{\bf \xi }^{\prime })},
\end{equation}
where the total projected Hamiltonian has the form

\begin{eqnarray}
H &=&\frac{e^{2}}{2\,}\int \int d{\bf x}d{\bf x}^{\prime }\,\,\Psi ^{*}({\bf %
x})\Psi ^{*}({\bf x}^{\prime })\,\frac{1}{\mid {\bf x}-{\bf x}^{\prime }\mid 
}\,\Psi ({\bf x}^{\prime })\Psi ({\bf x}) \\
&=&\frac{e^{2}}{2}\sum_{{\bf \alpha },{\bf \alpha }^{\prime }}\sum_{{\bf %
\beta },{\bf \beta }^{\prime }}\,M({\bf \alpha },{\bf \alpha }^{\prime }\mid 
{\bf \beta },{\bf \beta }^{\prime })\,\,\,\,a_{{\bf \alpha }}^{+}\,a_{{\bf %
\alpha }^{\prime }}^{+}\,a_{{\bf \beta }}\,a_{{\bf \beta }^{\prime }}\,, 
\nonumber
\end{eqnarray}
with the matrix element of the coulomb interaction given by 
\begin{equation}
M({\bf \alpha },{\bf \alpha }^{\prime }\mid {\bf \beta },{\bf \beta }%
^{\prime })\,=\int \int d{\bf x}d{\bf x}^{\prime }\,\,\varphi _{{\bf \alpha }%
}^{*}({\bf x})\varphi _{{\bf \alpha }^{\prime }}^{*}({\bf x}^{\prime })\,%
\frac{1}{\mid {\bf x}-{\bf x}^{\prime }\mid }\,\varphi _{{\bf \beta }%
^{\prime }}({\bf x}^{\prime })\varphi _{{\bf \beta }}({\bf x}).
\end{equation}
By using the anti-commutation relations $[a_{{\bf {q}}},a_{{\bf {q}^{\prime }%
}}^{+}]=\delta _{{\bf {q},{q}^{\prime }}}$, formula (11) can be expressed in
terms of our basis as,

\begin{equation}
E^{(2)}=\sum_{({\bf \eta ,\eta }^{\prime })}\sum_{({\bf \xi },{\bf \xi }%
^{\prime })}\frac{\mid e^{2}\int \int d{\bf x}d{\bf x}^{\prime }\,\Phi _{%
{\bf \eta },{\bf \eta }^{\prime }}^{*}({\bf x},{\bf x}^{\prime })\frac{1}{%
\mid {\bf x}-{\bf x}^{\prime }\mid }\Phi _{{\bf \xi },{\bf \xi }^{\prime }}(%
{\bf x},{\bf x}^{\prime })\mid ^{2}}{{\Large \epsilon }({\bf \eta })+{\Large %
\epsilon }({\bf \eta }^{\prime })-{\Large \epsilon }({\bf \xi })-{\Large %
\epsilon }({\bf \xi }^{\prime })},
\end{equation}
where the two particle states $\Phi $ are defined as 
\begin{equation}
\Phi _{{\bf \eta },{\bf \eta }^{\prime }}({\bf x},{\bf x}^{\prime })=\frac{%
\varphi _{{\bf \eta }}({\bf x})\varphi _{_{{\bf \eta }^{\prime }}}({\bf x}%
^{\prime })-\varphi _{_{{\bf \eta }^{\prime }}}({\bf x})\varphi _{{\bf \eta }%
}({\bf x}^{\prime })}{\sqrt{2}}.
\end{equation}
This quantity can be evaluated by use of the series (2), to obtain

\begin{eqnarray}
M({\bf \alpha }^{\prime },{\bf \alpha } &\mid &{\bf \beta },{\bf \beta }%
^{\prime })\,=N_{\phi _{0}}\delta _{P}^{(K)}({\bf \alpha }+{\bf \alpha }%
^{\prime },{\bf \beta }+{\bf \beta }^{\prime })\,I({\bf \beta }^{\prime }-%
{\bf \alpha },{\bf \beta }^{\prime }-{\bf \alpha }^{\prime })* \\
&&\frac{\varphi _{{\bf \alpha }}^{*}(-r_{o}^{2}{\bf n}\times (_{{\bf \beta }%
^{\prime }}-{\bf \alpha }))\varphi _{{\bf \alpha }^{\prime }}^{*}(r_{o}^{2}%
{\bf n}\times ({\bf \beta }^{\prime }-{\bf \alpha }))\,}{N_{{\bf \beta }%
^{\prime }}\,N_{{\bf \beta }}},  \nonumber
\end{eqnarray}
where the Coulomb interaction is contained in the function $I$, given by

\begin{eqnarray}
I({\bf \beta }^{\prime }-{\bf \alpha },{\bf \beta }^{\prime }-{\bf \alpha }%
^{\prime }) &=&\sum_{{\bf Q}}\frac{2\pi }{\mid {\bf \beta }^{\prime }-{\bf %
\alpha }+{\bf Q}\mid }\times  \nonumber \\
&&\exp \left( -\frac{r_{o}^{2}}{2}({\bf \beta }^{\prime }-{\bf \alpha }+{\bf %
Q})^{2}\right) \exp \left( i\;r_{o}^{2}\;{\bf n}\times ({\bf \alpha }%
^{\prime }-{\bf \beta }^{\prime })\cdot {\bf Q}\right) .
\end{eqnarray}
The inherent translation invariance of the problem is reflected in the delta
function expressing the conservation of the quasimomentum of the four
particle states defining the matrix element,

\begin{equation}
\delta _{{\bf P}}^{(K)}({\bf \eta }+{\bf \eta }^{\prime },{\bf \xi }+{\bf %
\xi }^{\prime })=\sum_{{\bf Q}}\delta ^{(K)}({\bf \eta }+{\bf \eta }^{\prime
},{\bf \xi }+{\bf \xi }^{\prime }+{\bf Q}).
\end{equation}
Finally, by using the above expression for the matrix element, the second
order correction to the energy per particle can be expressed as

\begin{equation}
{\Large \epsilon }^{(2)}=\frac{E^{(2)}}{N}=\frac{1}{2(2\pi )^{2}}\int \int
\int \frac{d{\bf q}}{(2\pi )^{2}}\frac{d{\bf q}^{\prime }}{(2\pi )^{2}}\frac{%
d{\bf p}}{(2\pi )^{2}}\;\frac{M({\bf q},{\bf q}^{\prime }\mid {\bf p})}{%
{\Large \epsilon }({\bf q})+{\Large \epsilon }({\bf q}^{\prime })-{\Large %
\epsilon }({\bf p})-{\Large \epsilon }({\bf q}+{\bf q}^{\prime }-{\bf p})}%
\frac{e^{2}}{\epsilon \text{ }r_{o}},
\end{equation}
where the momentum variables have been rescaled to be dimensionless through
the changes of variables ${\bf q}=a\ {\bf \eta },$ ${\bf q}^{\prime }=a\ 
{\bf \eta }^{\prime },$ ${\bf p}=a\ {\bf \xi }$, and the usual equivalence
in the high area limit $\sum_{{\bf q}}g({\bf p})\equiv S\int \frac{d{\bf q}}{%
(2\pi )^{2}}g({\bf p})$ has been employed. The expression for the kernel in
(13) becomes

\[
M({\bf q},{\bf q}^{\prime }\mid {\bf p})=\mid I({\bf q}^{\prime }-{\bf p},%
{\bf q}-{\bf p})\vartheta _{{\bf q}}(-\frac{{\bf n}\times ({\bf q}^{\prime }-%
{\bf p})}{2\pi })\vartheta _{{\bf q}^{\prime }}(\frac{{\bf n}\times ({\bf q}%
^{\prime }-{\bf p})}{2\pi })- 
\]

\begin{center}
\begin{equation}
-I({\bf p}-{\bf q},{\bf p}-{\bf q}^{\prime })\vartheta _{{\bf q}}(-\frac{%
{\bf n}\times ({\bf p}-{\bf q})}{2\pi })\vartheta _{{\bf q}^{\prime }}(\frac{%
{\bf n}\times ({\bf p}-{\bf q})}{2\pi })\mid ^{2}\frac{1}{\vartheta _{{\bf q}%
}(0)\vartheta _{{\bf q}^{\prime }}(0)\vartheta _{{\bf p}}(0)\vartheta _{{\bf %
q}+{\bf q}^{\prime }-{\bf p}}(0)}
\end{equation}
\end{center}

Relation (19) allows an estimate of $\epsilon ^{(2)}$ by evaluating the
three momentum integrals. The integration was performed partitioning the
Brillouin zone into a lattice of $(2n+1)^{2}$ points, over which the
integration variables take the values ${\bf q}(m_{1},m_{2})=\frac{\pi }{2n+1}
$ $(m_{1}-m_{2},m_{1}+m_{2})$, $-n\leq m_{1},m_{2}\leq n,$ and similarly for 
${\bf q}^{\prime }(m_{1},m_{2})$, ${\bf p}(m_{1},m_{2})$. These partitions
have the property that the points do not touch the boundary of $F$, and from
which they are at least a distance equal to half the distance between the
points of the partition. This property was implemented in order to avoid in
a regular manner the singularity which appears when the momenta of the two
states inside and outside $F$ are all on the Fermi boundary. When at least
one of the states is outside $F$ the difference of energies in (53) is
always non vanishing due to the form of the self energy surface. The results
in units of $\frac{e^{2}}{\varepsilon r_{o}}$ for partitions involving up to
225 points are

$
\begin{array}{cc}
n & \epsilon ^{(2)}\ (\frac{e^{2}}{r_{o}}) \\ 
1 & -0.0572742 \\ 
2 & -0.02460706 \\ 
3 & -0.02515106 \\ 
4 & -0.02762403 \\ 
5 & -0.02992157 \\ 
6 & -0.03175841 \\ 
7 & -0.0307658
\end{array}
$

\smallskip Comparison with the $\gamma =1$ value $0.002e^{2}/ \varepsilon
r_{o}$ shows an increase in the correction for $\gamma =1/2$ of over an
order of magnitude. Together with the significantly smaller space
fluctuations of the charge density, we interpret this result as a larger
sensitivity of the $\gamma =1/2 $ state to the introduction of correlations,
leading to a faster lowering of the energy and possibly melting of the
charge density wave. Extension of this work to other filling fractions will
be reported elsewhere.

\section{Summary}

An analytic solution of the Hartree-Fock problem at filling $\nu =\frac 12$
and half a particle per unit cell ($\gamma =\frac 12)$ has been discussed.
The same state was formerly studied numerically.\cite{claro2} The solution
is found to have a more uniform charge density in space than the $\gamma =1$
state of Yoshioka and Lee, and the correction to the energy is an order of
magnitude larger than that obtained with one electron per cell. Besides
yielding no gap as required for filling $1/2$, our results suggest that the
Hartree-Fock state with half of an electron per unit cell is a better
perturbative precursor to the true ground state of the system, and calls for
a more detailed investigation of its properties for the filling considered,
as well as other fractions.

It should be stressed that after finishing this work Dr. N. Maeda have
communicated us about two references:\cite{maeda1,maeda2}   in which the
same HF electronic state at $\nu =\frac 12\ $was  considered by him and
other  collaborators as possibly related with the composite fermions at this
filling fraction. In the present work, from a technical point of view,  we
have intended to present and analytical solution to the mean field problem
as a new result. Alternatively, on the physical side, our main purpose was
to decide about wether the enhanced electron overlapping at $\gamma =\frac 12
$ is able to also increase the correlation energy of the state at $\gamma =%
\frac 12$ over the one for $\gamma =1.$ 

\section{Acknowledgments}

This work was supported in part by Fondecyt 1990425 and a C\'{a}tedra
Presidencial en Ciencia (FC). The granting of the travel expenses by the
South-South Fellowship programme of the Third World Academy of Sciences
(TWAS) is also greatly acknowledged.The helpful support for two of the
authors (A.C.) and (F.C.) of the AS ICTP\ Associateship Programme is also
deeply appreciated.

\section{Apendix A}

The following properties of the basis functions under translations and
reflections are needed in the body of the text.

\subsection{Translations}

Let us argue that the effect of an arbitrary translation on the basis
functions is, modulo a phase factor, equivalent to a shift in the momentum
label\cite{ferrari}. Operating over a basis function twice with the
translation operator involving a lattice vector $\ell $ and an arbitrary
vector $a$ one gets,

\begin{eqnarray}
T_{{\bf a}}T_{{\bf \ell }}\ \varphi _{{\bf p}}({\bf x}) &=&\lambda _{{\bf p}%
}({\bf \ell })T_{{\bf a}}\varphi _{{\bf p}}({\bf x})  \eqnum{A.1} \\
&=&\exp (-2\frac{ie}{\hbar c}{\bf A}({\bf a}).{\bf \ell })\ T_{\ell }T_{{\bf %
a}}\ \varphi _{{\bf p}}({\bf x}),  \nonumber
\end{eqnarray}
where we have used Eq. (4) and the identity

\[
T_{{\bf a}}T_{{\bf \ell }}=\exp (2\frac{ie}{\hbar c}A({\bf a}).{\bf \ell }%
)T_{{\bf \ell }}T_{{\bf a}}\ . 
\]
Again using (4),

\[
T_{{\bf \ell }}T_{{\bf a}}\ \varphi _{{\bf p}}({\bf x})=\lambda _{{\bf p}+2%
\frac{ie}{\hbar c}{\bf A}({\bf a})}({\bf \ell })T_{{\bf a}}\ \varphi _{{\bf p%
}}({\bf x}). 
\]
Then, taking into account that the set of eigenvalues defines uniquely the
wavefunctions modulo a phase, it follows

\begin{equation}
T_{{\bf a}}\ \varphi _{{\bf p}}({\bf x})=F_{{\bf p}}({\bf a})\ \varphi _{%
{\bf p}+2\frac{ie}{\hbar c}{\bf A}({\bf a})}({\bf x}).  \eqnum{A.2}
\end{equation}
That is, a magnetic translation is equivalent to a shift in the
quasi-momentum. The factor $F,$ being a phase, satisfies 
\[
F_{{\bf p}}({\bf a})F_{{\bf p}}^{*}({\bf a})=1. 
\]
We now consider a special translation in the vector ${\bf a}=r_{o}^{2}\ {\bf %
n}\times {\bf \delta },$ with ${\bf \delta }=(-\frac{\pi }{a},\frac{\pi }{a}%
).$ Using Eq. A.2 one finds that it amounts to a shift in the momentum by
the quantity $2\frac{e}{\hbar c}{\bf A}({\bf a})={\bf \delta }.$ Then, the
magnetic translation in a vector $a$ transforms any momentum{\bf \ }${\bf p}%
\in F$ into a corresponding momentum ${\bf p}^{\delta }\in B-F$ through the
relation 
\begin{equation}
{\bf p}^{\delta }={\bf p}+{\bf \delta }  \eqnum{A.3}
\end{equation}
thus transforming all of $F$ in all $B-F.$

\subsection{Reflections}

Consider the axis $D$ in the plane determined by the vector

\begin{equation}
{\bf d}=(1,1),  \eqnum{A.4}
\end{equation}
and define the operation $R$ of a reflection on $D$ within the plane by

\begin{eqnarray}
{\bf x}^{R} &=&(x_{2},x_{1})=R\ {\bf x},  \eqnum{A.5} \\
{\bf p}^{R} &=&R\ {\bf p},  \nonumber
\end{eqnarray}
and its self-inverse property

\[
R=R^{-1}. 
\]
After acting with the operator $R$ over any function $f$ the following
property follows:

\begin{eqnarray*}
RT_{\ell }\ f({\bf x}) &=&R\exp (\frac{ie}{\hbar c}{\bf A}({\bf \ell }).{\bf %
x})\ f({\bf x}-{\bf \ell }) \\
&=&\exp (\frac{ie}{\hbar c}{\bf A}({\bf \ell }).R{\bf x})\ f(R{\bf x}-{\bf %
\ell }) \\
&=&\exp (\frac{ie}{\hbar c}{\bf A}(R{\bf \ell }).{\bf x})\ f(R({\bf x}-R{\bf %
\ell })) \\
&=&T_{R{\bf \ell }}\ R\;f({\bf x}),
\end{eqnarray*}
from which one gets

\begin{eqnarray}
R\varphi _{{\bf p}}({\bf x}) &=&\frac{1}{N_{{\bf p}}}\sum_{{\bf \ell }%
}\left( -1\right) ^{\ell _{1}\ell _{2}}\exp (i\ R{\bf p}.{\bf \ell })\ T_{%
{\bf \ell }}\ \phi ({\bf x})  \eqnum{A.6} \\
&=&\varphi _{R{\bf p}}({\bf x}),  \nonumber
\end{eqnarray}
meaning that the action of a space reflection on the basis function is
equivalent to a reflection of its quasi-momentum.

\section{Apendix B}

That the lowest energy state for half filling is obtained by occupying all
states within the square bounded by the lines $ak_{x}+ak_{y}=\pi $ may be
shown as follows. First, let us inspect the commutation properties of the
reflection $R$ under the line containing the vector{\bf \ }${\bf d}=(1,1)$,
as well as the particular translation $T_{a}$ which transforms the filled
states $F$ into the empty sates in $B-F$, with the Hartree-Fock single
particle Hamiltonian. Both operations are defined in Appendix A. As the
action of this hamiltonian is represented by a kernel with two arguments due
to the projection on the first Landau level, it turns out to be useful to
employ, when appropriate, the Dirac notation. Then the single particle HF
equation can be expressed as

\begin{equation}
H_{U}^{(HF)}\mid \varphi _{{\bf p}}\rangle ={\Large \epsilon }({\bf p})\mid
\varphi _{{\bf p}}\rangle  \eqnum{B.1}
\end{equation}
where the kernel associated to the Hartree-Fock Hamiltonian is

\begin{eqnarray}
H_{U}^{(HF)}({\bf x},{\bf x}^{^{\prime }}) &=&\hbar \omega _{o}\delta ({\bf x%
},{\bf x}^{\prime })+\int d{\bf y}d{\bf y}^{\prime }\ P_{o}({\bf x},{\bf y})%
\frac{e^{2}}{\mid {\bf y}-{\bf y}^{\prime }\mid }\times  \eqnum{B.2} \\
&&{\Large \{(}\sum_{{\bf q}\in U}\varphi _{{\bf q}}^{*}({\bf y}^{\prime
})\varphi _{{\bf q}}({\bf y}^{\prime })-n_{o}{\Large )}\ P_{o}({\bf y},{\bf x%
}^{\prime })-  \nonumber \\
&&{\Large (}\sum_{{\bf q}\in U}\varphi _{{\bf q}}^{*}({\bf y}^{\prime
})\varphi _{{\bf q}}({\bf y}){\Large )}\ P_{o}({\bf y}^{\prime },{\bf x}%
^{\prime }){\Large \}.}  \nonumber
\end{eqnarray}
Here $P_{o}$ is the projection operator in the first Landau level and the
direct and exchange contributions are determined by the set $U$ formed by
the momenta associated to the selected filled states. At this point, let us
assume that $U=F$. From the properties of the reflections it follows that

\begin{equation}
R\int H_{F}^{(HF)}({\bf x},{\bf x}^{\prime })\ f({\bf x}^{\prime })\ d{\bf x}%
^{\prime }=\int H_{F}^{(HF)}(R{\bf x},R{\bf x}^{\prime })R\ f({\bf x}%
^{\prime })\ d{\bf x}^{\prime }.  \eqnum{B.3}
\end{equation}
Noting the invariance property that follows from Eq. A.8,

\begin{eqnarray*}
P_{o}(R{\bf x},R{\bf x}^{\prime }) &=&\sum_{{\bf q}\in B}\varphi _{{\bf q}}(R%
{\bf y})\ \varphi _{{\bf q}}^{*}(R{\bf y}^{\prime }) \\
&=&\sum_{{\bf q}\in B}\varphi _{R{\bf q}}({\bf y})\ \varphi _{R{\bf q}}^{*}(%
{\bf y}^{\prime }) \\
&=&P_{o}({\bf x},{\bf x}^{\prime })
\end{eqnarray*}
the invariance of the Hamiltonian follows,

\begin{eqnarray}
H_F^{(HF)}(R{\bf x},R{\bf x}^{\prime }) &=&\hbar \omega _o\delta ({\bf x},%
{\bf x}^{\prime })+\int d{\bf y}d{\bf y}^{\prime }\ P_o(R{\bf x},R{\bf y})%
\frac{e^2}{\mid R{\bf y}-R{\bf y}^{\prime }\mid }\times  \eqnum{B.4} \\
&&{\Large \{(}\sum_{{\bf q}\in F}\varphi _{{\bf q}}^{*}(R{\bf y}^{\prime
})\varphi _{{\bf q}}(R{\bf y}^{\prime })-n_o{\Large )}\ P_o(R{\bf y},R{\bf x}%
^{\prime })-  \nonumber \\
&&{\Large (}\sum_{{\bf q}\in F}\varphi _{{\bf q}}^{*}(R{\bf y}^{\prime
})\varphi _{{\bf q}}(R{\bf y}){\Large )}\ P_o(R{\bf y}^{\prime },R{\bf x}%
^{\prime }){\Large \}}  \nonumber \\
&=&H_F^{(HF)}({\bf x},{\bf x}^{\prime }),  \nonumber
\end{eqnarray}
where in the sums again use of A.8 has been made, as well as the fact that
if $q\in F$ then its reflection about $D$ is also in $F$. Using the above
property in Eq. B.3 one gets 
\[
R\int H_F^{(HF)}({\bf x},{\bf x}^{\prime })\ f({\bf x}^{\prime })\ d{\bf x}%
^{\prime }=\int H_F^{(HF)}({\bf x},{\bf x}^{\prime })R\ f({\bf x}^{\prime })d%
{\bf x}^{\prime } 
\]
>From the arbitrariness of $f$ then follows the commutativity of the
reflection operator with $H^{(HF)},$

\[
\lbrack H_F^{(HF)},R]=0. 
\]
and consequently

\begin{eqnarray*}
RH_{F}^{(HF)} &\mid &\varphi _{{\bf p}}\ \rangle ={\Large \epsilon }({\bf p}%
)\ R\mid \varphi _{{\bf p}}\ \rangle \\
&=&H_{F}^{(HF)}\ R\mid \varphi _{{\bf p}}\ \rangle \\
&=&{\Large \epsilon }(R\,{\bf p})\ \mid \varphi _{R{\bf p}}\ \rangle ,
\end{eqnarray*}
leading to the symmetry of the spectrum with respect to reflection about the
line $D$ defined by ${\bf d}=(1,1),$

\begin{equation}
{\Large \epsilon }({\bf p})={\Large \epsilon }(R\,{\bf p}).\   \eqnum{B.5}
\end{equation}
Clearly, the same conclusion arises with respect to reflections on the line $%
\overline{D\text{ }}$defined by $\overline{{\bf d}}=(-1,1).$

The next task is to consider symmetry properties under a translation in the
special vector $a$ defined in Appendix A . In this sense, it is worth noting
that if ${\bf p}\in F$ then ${\bf p}+{\bf \delta }\in B-F$ and moreover, $R\ 
{\bf p}$ exactly gives{\Large \ }${\bf p}+{\bf \delta }$ under a reflection
about the line defined by the boundary of $F$ and $B-F. $ As we shall see,
the relation between values of the self energies under such operations will
allow us to show the property we are after.

We start out by considering the states in the sector $B-F$ and exploit
particle hole symmetry. Adding the hamiltonian obtained through such states
to that given in B.4 one obtains the Hartree-Fock form appropriate for a
filled Brillouin zone, 
\begin{eqnarray}
H_{B}^{(HF)}({\bf x},{\bf x}^{^{\prime }}) &=&H_{F}^{(HF)}({\bf x},{\bf x}%
^{^{\prime }})+H_{B-F}^{(HF)}({\bf x},{\bf x}^{^{\prime }})  \eqnum{B.6} \\
&=&\hbar \omega _{o}\delta ({\bf x},{\bf x}^{\prime })-\int d{\bf y}d{\bf y}%
^{\prime }\ P_{o}({\bf x},{\bf y})\frac{e^{2}}{\mid {\bf y}-{\bf y}^{\prime
}\mid }  \nonumber \\
&&\sum_{{\bf q}\in B}\varphi _{{\bf q}}^{*}({\bf y}^{\prime })\varphi _{{\bf %
q}}({\bf y})\,\ P_{o}({\bf y}^{\prime },{\bf x}^{\prime })  \nonumber \\
&=&\hbar \omega _{o}\delta ({\bf x},{\bf x}^{\prime })-\int d{\bf y}d{\bf y}%
^{\prime }\ P_{o}({\bf x},{\bf y})\frac{e^{2}}{\mid {\bf y}-{\bf y}^{\prime
}\mid }  \nonumber \\
&&{\Large \ }P_{o}({\bf y},{\bf y}^{\prime })\ P_{o}({\bf y}^{\prime },{\bf x%
}^{\prime }).  \nonumber
\end{eqnarray}
It can be readily verified that the kernel $H_{B}^{(HF)}$commutes with all
the magnetic translations, that is, 
\begin{equation}
\lbrack T_{a},H_{B}^{(HF)}]=0.  \eqnum{B.7}
\end{equation}
Henceforth, since in particular it commutes with all the $T_{{\bf \ell }}$
leaving invariant the lattice $L,\ $it follows that the $\varphi _{{\bf p}}$
are eigenfunctions of $H_{B}^{(HF)}$. Additionally, they all have exactly
the same eigenvalue. This can be verified by considering a translation in
any vector $a^{\prime }=r_{o}^{2}n\times \delta ^{\prime }$ commuting with
it, which changes the momenta in an also arbitrary quantity ${\bf \delta }%
^{\prime }$ (Appendix A). One has $H_{B}^{(HF)}\mid \varphi _{p}\ \rangle
=\epsilon _{B}({\bf p})\mid \varphi _{{\bf p}}\ \rangle ,$that may also be
written as $H_{B}^{(HF)}T_{r_{o}^{2}{\bf n}\times {\bf \delta }^{\prime
}}\mid \varphi _{{\bf p}-{\bf \delta }^{\prime }}\ \rangle =\epsilon _{B}(%
{\bf p})T_{r_{o}^{2}{\bf n}\times {\bf \delta }^{\prime }}\mid \varphi _{%
{\bf p}-{\bf \delta }^{\prime }}\ \rangle $. Multiplying this latter
expression from the left by the inverse translation one gets $%
H_{B}^{(HF)}\mid \varphi _{{\bf p}-{\bf \delta }^{\prime }}\ \rangle
=\epsilon _{B}({\bf p})\mid \varphi _{{\bf p}-{\bf \delta }^{\prime }}\
\rangle $, which also equals $\epsilon _{B}({\bf p}-{\bf \delta }^{\prime
})\mid \varphi _{{\bf p}-{\bf \delta }^{\prime }}\ \rangle $. Therefore, $%
\epsilon _{B}({\bf p})=\epsilon _{B}({\bf p}-{\bf \delta }^{\prime
})=\epsilon _{B},$ a constant. This property enforces the proportionality of 
$H_{B}^{(HF)}$ with the projection operator $P_{o}$. At this point it is
worthwhile to recall that since we are considering functions on the first
Landau level, the Dirac delta function appearing in the definition of the HF
kernels is equivalent to the projection operator. Then, 
\begin{equation}
H_{B}^{(HF)}({\bf x},{\bf x}^{^{\prime }})=\hbar \omega _{o}\delta ({\bf x},%
{\bf x}^{\prime })+{\Large \epsilon }_{L}\,P_{o}({\bf x},{\bf x}^{\prime }) 
\eqnum{B.8}
\end{equation}
Thus, all the basis functions are eigenfunctions of the Hartree-Fock
hamiltonian associated with the filled Landau level $H_{B}^{(HF)}\mid
\varphi _{{\bf p}}\ \rangle =\epsilon _{B}\mid \varphi _{{\bf p}}\ \rangle $%
, with the same eigenvalue $\epsilon _{B}=\hbar \omega _{o}+\epsilon _{L}.$

The next point to consider is the connection between the kernels $%
H_{F}^{(HF)}$ and $H_{B-F}^{(HF)}.$ It follows from expressing the basis
functions with momenta in $B-F$ by means of translations of the basis states
with momenta in $F,$ and considering the transformation properties under the
special translations $T_{r_{o}^{2}\ {\bf n}\times {\bf \delta }}$ of the
projection operator and the Coulomb potential, as discussed in Appendix A.
It follows

\begin{eqnarray}
H_{B-F}^{(HF)}({\bf x},{\bf x}^{\prime }) &=&\hbar \omega _{o}\delta ({\bf x}%
,{\bf x}^{\prime })+\int d{\bf y}d{\bf y}^{\prime }\ P_{o}({\bf x},{\bf y})%
\frac{e^{2}}{\mid {\bf y}-{\bf y}^{\prime }\mid }  \eqnum{B.9} \\
&&{\Large \{}\sum_{{\bf q}\in F}T_{r_{o}^{2}\ {\bf n}\times {\bf \delta }%
}\varphi _{{\bf q}}({\bf y}^{\prime })\text{ }[T_{r_{o}^{2}\ {\bf n}\times 
{\bf \delta }}\text{ }\varphi _{{\bf q}}({\bf y}^{\prime })]^{*}-n_{o}%
{\Large )}\ P_{o}({\bf y},{\bf x}^{\prime })-  \nonumber \\
&&\sum_{{\bf q}\in F}T_{r_{o}^{2}\ {\bf n}\times {\bf \delta }}\,\varphi _{%
{\bf q}}({\bf y})\,[T_{r_{o}^{2}\ {\bf n}\times {\bf \delta }}\varphi _{{\bf %
q}}({\bf y}^{\prime })\,]^{*}{\Large \ }P_{o}({\bf y}^{\prime },{\bf x}%
^{\prime }){\Large \}}  \nonumber \\
&=&\hbar \omega _{o}\delta ({\bf x},{\bf x}^{\prime })+\int d{\bf y}d{\bf y}%
^{\prime }\ P_{o}({\bf x},{\bf y})T_{r_{o}^{2}\ {\bf n}\times {\bf \delta }%
}^{({\bf y})}\frac{e^{2}}{\mid {\bf y}-{\bf y}^{\prime }\mid }  \nonumber \\
&&{\Large \{(}\sum_{{\bf q}\in F}\varphi _{{\bf q}}^{*}({\bf y}^{\prime
})\varphi _{q}({\bf y}^{\prime })-n_{o}{\Large )}\ T_{-r_{o}^{2}\ {\bf n}%
\times {\bf \delta }}^{({\bf y})}P_{o}({\bf y},{\bf x}^{\prime })-  \nonumber
\\
&&\sum_{{\bf q}\in F}\varphi _{{\bf q}}({\bf y})\,\varphi _{{\bf q}}^{*}(%
{\bf y}^{\prime })\,\ T_{-r_{o}^{2}\ {\bf n}\times {\bf \delta }}^{({\bf y}%
^{\prime })}P_{o}({\bf y}^{\prime },{\bf x}^{\prime }){\Large \},}  \nonumber
\end{eqnarray}
which may be expressed in the more compact Dirac notation as,

\begin{equation}
H_{B-F}^{(HF)}=T_{r_{o}^{2}\ {\bf n}\times {\bf \delta }%
}\,H_{F}^{(HF)}T_{-r_{o}^{2}\ {\bf n}\times {\bf \delta }}.  \eqnum{B.10}
\end{equation}
Physically, this relation means that the HF state associated to $F$ is
simply the space translation of the state associated to $B-F.\ $Therefore,
within the first Landau level the following relation holds, 
\begin{equation}
{\Large \epsilon }_{B}=H_{F}^{(HF)}+T_{r_{o}^{2}\ {\bf n}\times {\bf \delta }%
}\,H_{F}^{(HF)}T_{-r_{o}^{2}\ {\bf n}\times {\bf \delta }},  \eqnum{B.11}
\end{equation}
which after acting over a basis function produces

\begin{equation}
{\Large \epsilon }_{B}={\Large \epsilon }({\bf p})+{\Large \epsilon }({\bf p}%
-{\bf \delta }),  \eqnum{B.12}
\end{equation}
The sum of the energies at$\ p\ $and the shifted value is strictly constant.
Using the symmetry of the spectrum under reflection it also follows that 
\begin{equation}
{\Large \epsilon }_{B}={\Large \epsilon }({\bf p})+{\Large \epsilon }(R({\bf %
p}-{\bf \delta })).  \eqnum{B.13}
\end{equation}
Thus, the sum of the energies associated to the states which have momenta
related by a reflection in the boundary of the regions $F$ and $B-F$ is also
exactly constant. Finally by selecting ${\bf p}={\bf p}_{F}$ , with ${\bf p}%
_{F}$ in the boundary between both regions, and noticing that in that case $%
{\bf p}_{F}=R({\bf p}_{F}-{\bf \delta }),$ it follows that, due to the
continuity of the spectrum (absence of gaps) the energy at any point of the
boundary of the filled states strictly equals $\epsilon ({\bf p}%
_{F})=\epsilon _{F}/2.$ Therefore, the criterium for a minimum given in
reference\cite{cabo2} is satisfied, and the Hartree-Fock solution with all
states in $F$ has a local minimum of the energy.

\end{document}